# Rietveld refinement and electron density distribution in $Cu_xV_2O_5$

R. Nithya, Sharat Chandra[*], G.L.N. Reddy[1], H.K. Sahu and V. Sankara Sastry,

*Materials Science Division, Indira Gandhi Centre for Atomic Research, Kalpakkam 603102, Tamil Nadu, India.*
[1]*Present Address: Centre for Compositional Characterization of Materials, ECIL Post, Hyderabad 500 062, Andhra Pradesh, India*

**Abstract**

Room temperature powder x-ray diffraction measurements have been carried out on polycrystalline samples of $Cu_xV_2O_5$ with x = 0.33, 0.40, 0.50, 0.55 and 0.60, prepared by solid state reaction in evacuated quartz tubes. All the samples were found to be in single-phase monoclinic structure. Rietveld full profile refinement of the x-ray diffraction data yielded the lattice parameters and fractional atom coordinates. Maximum entropy inversion of the diffraction data was done in order to obtain the variation in the spatial electron density distribution as a function of copper composition. These density maps show progressive positional and charge disorders in the V-O network induced by addition of copper. It is shown that the electrons donated by copper atoms are transferred to the specific vanadium atom sites.

## 1. Introduction

Vanadium combines with many other elements and forms a large variety of compounds and alloys such as chemical compounds, intermetallics, substitutional alloys and interstitial alloys depending on the bonding characteristics and structure. In these materials, vanadium exhibits different oxidation states ranging from +2 to +5 and this governs the different physical properties and chemistry to a large extent. In the present work, we study the changes in the structure and electron redistribution in the copper-vanadium bronzes with different copper content.

Pure $V_2O_5$ is a stable compound where vanadium is in its highest oxidation state of +5 with $3d^0$ electronic configuration. In $V_2O_5$, vanadium has five oxygen coordination generating $VO_6$ octahedrons. These octahedrons share their corners with adjacent $VO_6$ and form double chains along the b-axis. The chains are linked to form the a-b layers and these layers are stacked along the c-direction. Thus, $V_2O_5$ forms a tunnel structure with large interstitial spaces [1]. Incorporation of $M^+$ cations introduces electrons in the $V_2O_5$ host lattice leading to the formation of $V^{4+}$ ions. Depending on the element, M, and its concentration, x, in $M_xV_2O_5$, several phases of vanadium bronzes have been reported and are known to exhibit typical physical and chemical properties [2]. $\beta$-$M_xV_2O_5$ bronze has a variable extent of intercalation (x=0.17 to 0.67) depending on element, M. It inherits the tunnel structure of the parent $V_2O_5$ compound and is one of the best conductors among all the vanadium bronzes, different aspects of which have been investigated [3-5]. Many of the compounds in the β-phase are known to exhibit abrupt changes in property at x~0.33. The physical properties of these phases have been interpreted based on their crystal structure which is similar to the structure of $Na_{2-x}V_6O_{15}$ (space group A2/m) as determined by Wadsley [1] using Patterson maps and electron density projections. Wadsley showed that there is only one site for Na, three nonequivalent sites for vanadium and eight nonequivalent sites for oxygen atoms. The three vanadium atoms, V1, V2 and V3 form three types of strongly distorted $VO_6$ and $VO_5$ polyhedra that are linked by common oxygen ions. There are two sites, labeled $M_1$ and $M_1$', which are randomly occupied by $M^+$ ions [6]. However, within a tunnel the $M_1$-$M_1$' separation in ac plane is only 0.195 nm. As a result, simultaneous occupancy of both the sites by bigger ions like Na, K etc., is inhibited. Thus, in such cases, the maximum value for x in $\beta$-$M_xV_2O_5$ with larger cations is limited to ~0.33. In compounds having smaller cations, such as $Cu^+$ and $Li^+$, single-phase regions extend up to x=0.64 and 0.49 respectively. This suggests that both the $M_1$ and $M_1$' sites within the tunnels accommodate the cations [6,7]. While the $Li^+$ compounds exhibit both the β and β' phases, the crystal structure of copper vanadium bronzes exhibits only the $\beta^{'}$–phase [7].

---

[*]Corresponding author email: sharat@igcar.ernet.in

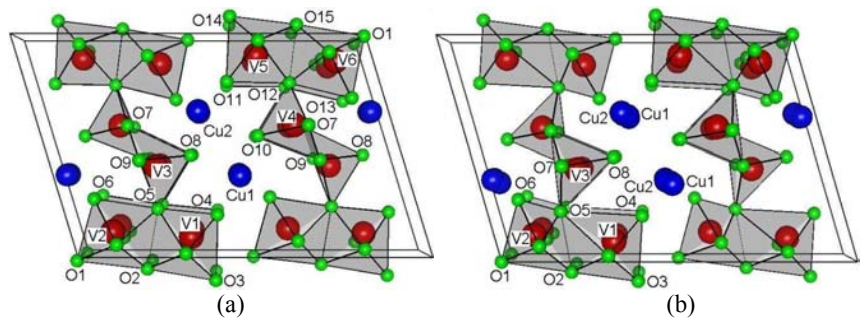

Fig.1. Unit cell of β'-$Cu_xV_2O_5$ in (010) orientation: (a) Cm and (b) C2/m. The octahedrons and bipyramids formed by V and O atoms are shown. Atoms in asymmetric unit cell have been labeled. Both the models have been used for Rietveld refinement.

Copper vanadium bronzes exist over a wider range of copper content as compared to other analogous compounds; thus wider range of electron density in the unit cell. Also, they exhibit no discontinuities in any of the physical properties as a function of x. The ability to accommodate larger number of metal ions without any changes in the crystal structure makes this compound an ideal candidate for studying the physical properties as a function of electron concentration. Two structural models of copper vanadium bronzes based on Cm (space group 8, setting 1) and C2/m (space group 12, setting 1) have been reported in literature [8,9]. In the Cm representation [8], there are two crystallographically nonequivalent sites for copper (Wyckoff multiplicity 4b, general position x,y,z), six sites for vanadium (Wyckoff multiplicity 2a, general position x,0,z) and 15 sites for oxygen (Wyckoff multiplicity 2a, general position x,0,z). In the centrosymmetric C2/m representation [9], there are two nonequivalent sites for copper (Wyckoff multiplicity 4i for Cu1 and 8j for Cu2, general position x,0,z and x,y,z respectively), three nonequivalent sites for vanadium (Wyckoff multiplicity 4i, general position x,0,z) and eight nonequivalent sites for oxygen (Wyckoff multiplicity 2a for O1 and 4i for O2 to O8 with the corresponding general positions 0,0,0 and x,0,z). In addition to Galy et al [8] and Savariault et al [9], Kato et al have also studied the single crystal specimens of the copper vanadium bronzes [10]. Crystal structure of β'-$Cu_xV_2O_5$ unit cell in (010) orientation is shown in Fig.1 for both the spacegroups. It can be seen that both the models yield essentially the same crystal structure with the difference being in the copper atom positions. The unit cell contains six $V_2O_5$ formula units per unit cell in both the cases. All the possible sites for the copper ions are located between $VO_6$ octahedra and are gradually filled with increasing x value. The Cu atoms donate the electrons to the V-O sublattice and do not influence any of the observed physical properties. Thus the behavior of the electrons in the V-O sublattice determines the observed physical properties.

Casalot et al., [5] reported their studies on the transport properties of polycrystalline β'-$Cu_xV_2O_5$ and showed the associated activation energy to decrease from 0.08eV at x=0.3 to about 0.02 eV at x=0.6. Mori et al., [11] reported a quasi-one dimensional metallic electrical conductivity in oxygen deficient, copper rich β'-$Cu_{0.52}V_2O_{4.8}$ down to 1.5 K. Presence of metallic conductivity at such a low temperature has been ascribed to the electron transport within the linear chain of the delocalized orbitals of V3. Electron paramagnetic resonance measurements [12] performed on single crystals of β'-$Cu_xV_2O_5$ with x between 0.3 and 0.6 showed only the existence of monovalent copper ions. On the other hand, Savariault et al [9] reported the existence of both the $Cu^+$ and $Cu^{2+}$ species in the copper compounds using X-ray photoelectron spectroscopy [9]. From diffuse x-ray scattering studies on β-vanadium bronzes, Kanai et al., [2] found structural changes, which were attributed to a possible bipolaron ordering. Chakraverty et al, [3] observed a rather large linear electronic coefficient in the low temperature specific heat of $M_xV_2O_5$ in the β-phase. This was ascribed to the bipolaron formation of $V^{4+}$-$V^{4+}$ pairs. A report of spin-peierls transition in α-$NaV_2O_5$ [13] has triggered a rejuvenated interest in the vanadium bronzes. Yamada and Ueda [14] reported phase transitions in β'-$Cu_xV_2O_5$ system for the first time from transport, magnetic and structural studies at different temperatures. Thus, even though the vanadium bronzes have been studied for several decades, the new observation of such phenomena has lead to renewed interest in the study of exotic phase transitions involving spin, charge and lattice in β'-$Cu_xV_2O_5$ systems. It is therefore expected that these phenomena should be guided by the electronic density distribution in the material because many of these properties are related to the mechanism of localization of the electrons donated by copper (to one of the vanadium atoms). We have undertaken this study of mapping the electron density distributions in the range $0.33 \leq x \leq 0.60$ at%, in order to identify the specific vanadium sites where the donated electrons get localized. We use the Maximum Entropy Method (MEM) to obtain the electron density maps by inversion of the diffraction data.

MEM has been used in many fields including crystallography, as reviewed by Gilmore [15]. It is a non-linear method that has its roots in information and probability theory, and is originally designed to reconstruct the most probable and least biased probability distribution in an underdetermined situation. In crystallography it is often used to calculate strictly positive electron densities using single crystal or powder diffraction data. Starting with noisy and underrepresented image data, the traditional practice is to optimally select a value of the regularization parameter that makes the $\chi^2$ misfit statistics equal to the number of observations. This procedure is ad hoc and does not allow for the reduction in the effective number of the degrees of freedom caused by fitting accurate data. On the other hand, Bayesian determination of the regularization parameter allows for an accurate estimation of the overall missing data, thus yielding a higher resolution in the structure of the final image as compared to the traditional methods [16]. Thus the MEM density is less affected by missing reflections and is devoid of the series termination errors associated with the Fourier inverted density. Since not all phase factors can be determined unambiguously and because many reflections often overlap in the powder pattern, the dataset used to calculate the electron density map from powder data is essentially incomplete as compared with the single crystal case with well-separated reflections. This apparent deficiency does not pose a problem in MEM where all available information in the experimental data, i.e. phased, unphased, ambiguously phased and overlapping reflections can be used in the calculation of the charge density. On the other hand, in the Fourier inversion only the phased reflections can be used, often leading to severe artificial distortions in the calculated density [17]. Electron density maps can provide clues to the various possible electronic interactions between different atoms in the unit cell. Takata et al, [18] first showed the difference between the nature of bonding in Silicon (covalent) and LiF (ionic) by analyzing the electron density distribution obtained using MEM and reproduced accurate x-ray structure factors for the various x-ray reflections. Following this study, many research groups have been using MEM formalism to get quantitative electron and nuclear density distributions in crystalline compounds [18-25].

The present work involves preparation of single-phase polycrystalline samples of $\beta'$-$Cu_xV_2O_5$ (x = 0.33 to 0.60) and characterization of their structure by room temperature powder X-ray diffraction technique. The collected XRD data have been analyzed using the Rietveld analysis. The data have also been inverted for obtaining precise electron density distribution maps through an iterative procedure based on maximum entropy method. The electron density maps thus obtained help to visualize the redistribution and localization of electrons donated by copper to $VO_6$ network and to the lattice.

## 2. Experimental

Polycrystalline samples of $Cu_xV_2O_5$ (x=0.33, 0.40, 0.50, 0.55 and 0.60) were prepared using 99.99% pure CuO, $V_2O_5$ and $V_2O_3$ in stoichiometric amounts. The mixture was palletized after thorough grinding for homogenization, sealed in quartz tubes under vacuum and heat treated at 620º C for 20 hours followed by quenching into ice water. After obtaining the sample pellets, the pellets were powdered again by thorough grinding and the powder was sprinkled onto the Silicon (911) substrate holder for x-ray diffraction (XRD) measurements. XRD data were collected using a STOE diffractometer in the Bragg-Brentano geometry with Cu $K_\alpha$ radiation in the angular range $10º \leq 2\theta \leq 120º$ in step scan mode with counting time of 15 seconds per step and $2\theta$ steps of 0.02º. PowderX developed by Cheng Dong [26] was used to subtract the background from the raw data. Rietveld refinement using RIETAN-2000 package [27] was carried out on the background corrected data. After the final convergences were achieved, the structure factors were extracted from the fitted data and used as input for the MEM calculations. PRIMA developed by Dilanian and Izumi [28] to invert X-ray and neutron diffraction data by maximum entropy method was used to calculate electron density maps for samples with different copper compositions. PRIMA also generates a file containing the structure factors calculated after the MEM cycle convergence is achieved and this is then used to further refine only the structure factors in RIETAN-2000. These refined structure factors are then again used as input for PRIMA. Whole-pattern fitting and MEM analysis are alternately repeated in this iterative procedure called REMEDY cycles, whereby the bias imposed by a structural model upon final electron densities is minimized. Electron density distribution changes noticeably with decrease in final R-parameters during the REMEDY cycles, which is a strong piece of evidence for the reduction in the bias of the structural model [29]. This method is capable of modeling chemical bonds, nonlocalized electrons, and anharmonic thermal motion more adequately than the conventional Rietveld method as has been shown by Claridge et al for complex fulleride superstructure-decoupling in $Ca_{3+x}C_{60}$ [30].

## 3. Results and Discussion

The experimental x-ray powder diffraction patterns for x = 0.33, 0.40, 0.50, 0.55, and 0.60 are shown in Fig.2 after background correction in the $2\theta = 10º - 70º$ range for clarity. All samples were found to be in single phase. The starting values of lattice parameters were taken from Savariault et al [9] then refined by RIETAN-2000 in the full $2\theta$ range of $10º - 120º$. Whole powder pattern fitting of x-ray diffraction data was carried out for both the structural models based on Cm and C2/m shown in Fig.1. The cell parameters and fractional atom coordinates given by Galy et al. for space group Cm [8] and Savariault et al for space group C2/m [9] were used as starting input parameters and were allowed to relax during the iterations. The model proposed by Galy et al includes 23 atoms in the asymmetric unit cell, whereas the model proposed by Savariault et al contains 13 atoms in the asymmetric unit cell. The background was fitted using sixth order Legendre polynomial. While the absorption correction is not strictly needed in Bragg-Brentano-type x-ray powder diffraction geometry using flat-plate samples [28], the zero shift of the system was chosen as one of the fitting parameters. Split pseudo-Voigt function for relaxed reflections was used to fit the diffraction profiles [28]. This function was required to fit properly the numerous overlapping reflections observed in the data. During refinement, the parameter turn-on procedure suggested by Young [31] was largely used to select the parameters to refine. The cell parameters were refined in the following order a,b,c and β, followed by the fractional atom coordinates. In the initial cycles only the parameters corresponding to a single atom were refined one by one. When the system was near convergence, all the cell and atomic parameters (a, b, c, β, x, y, z, occupancy for Cu and thermal parameters) were refined simultaneously to minimize the correlations between the parameters of different atoms. This was followed by one cycle of refinement on the shift, background and profile parameters and then by one more cycle of simultaneous refinement on the cell and atom parameters. In the last cycle all the parameters were varied simultaneously to get the estimated variances of the various parameters.

| x | $R_{wP}$, $R_P$, S & Dwd | a, b, c (in nm), β (in degrees) | Cu (g) |
|---|---|---|---|
| 0.33 | 9.29,6.91,1.17,1.79 | 1.5252(1),0.3620(1),1.0111(1),106.973(6) | 0.240(7) |
| 0.40 | 8.31,5.84,1.17,1.29 | 1.5249(1),0.3622(1),1.0106(1),106.736(7) | 0.320(7) |
| 0.50 | 9.03,6.32,1.17,1.62 | 1.5228(1),0.3622(1),1.0096(1),106.570(7) | 0.360(7) |
| 0.55 | 9.05,6.31,1.17,1.69 | 1.5215(1),0.3624(1),1.0090(1),106.382(7) | 0.390(7) |
| 0.60 | 8.36,6.08,1.17,1.69 | 1.5213(1),0.3633(1),1.0099(1),106.133(8) | 0.460(7) |

Table.I: The final $R_{wP}$, $R_P$ and S parameters, Durbin Watson Statistics (Dwd), refined cell parameters and Cu atom occupancy (g) for various Cu concentrations in $Cu_xV_2O_5$. Occupancy for all other atoms is 1.00.

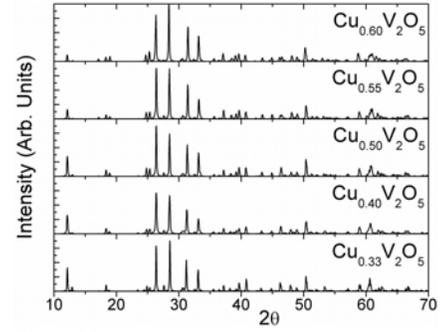

Fig.2. X-ray powder diffraction patterns for Cu = 33, 40, 50, 55 and 60 at%.

The thermal parameters were constrained to be equal for similar atomic species, but were allowed to vary independently for different Cu compositions. During the refinement it was found that the preferred orientation effects were practically absent. Both the Cm and C2/m spacegroups with the corresponding number of atoms quoted in literature gave the same order of R values and goodness of fit (e.g., for x = 0.40, Cm representation yields $R_{wP}$ = 17.47, $R_P$ = 13.58, S = 1.1272, Durbin Watson Statistics = 1.6971 and the C2/m representation yields $R_{wP}$ = 21.61, $R_P$ = 16.57, S = 1.3943, Durbin Watson Statistics = 1.2521). A further decrease in the R parameters was not possible using these unit cell models. The model based on Cm spacegroup contains 23 atoms in the asymmetric unit cell and thus an even larger number of total refinable parameters, making proper convergences difficult to achieve. While in the C2/m model, the number of atoms in the asymmetric unit cell is 13, it was found that it lead to very small bond lengths between the two Cu atoms and very large bond lengths between Cu and O atoms after few cycles of refinement, making the model clearly unphysical. It was seen that the x and z fractional atom coordinates for the Cu atom in position with Wyckoff multiplicity 4i were almost becoming equal to those for the second Cu atom in position with Wyckoff multiplicity 8j and the y fractional atom position for the second Cu atom was becoming of the order of $1\times10^{-2}$. Hence, it was decided to assign only one position to Cu atom in the specified C2/m unit cell model with Wyckoff multiplicity 8j. This is not in contradiction to the model presented in [9] where the authors have described all the possible positions that the Cu atom can take in the unit cell, but physically, the Cu atom can go to only one position out of all the possible sites. This alteration in the unit cell model makes both the Cm and C2/m representations equivalent, because with this change Cm unit cell can be obtained by a simple transformation of the C2/m unit cell. The transformation rules used for the purpose are discussed here: Cu (C2/m, x,y,z) generates two atoms in Cm (Wyckoff multiplicity 4b, Cu1, x,y,z and Cu2, 1-x,y,1-z), O1 (C2/m) generates one atom in Cm (Wyckoff multiplicity 2a, O1, 0,0,0) and atoms V1-V3 and O2-O8 (C2/m) generate two atoms each in Cm (Wyckoff multiplicity 2a, x,0,z and 1-x,0,1-z). Further, this choice of a single position for the Cu atom leads to a dramatic improvement in the goodness of fit and R parameters. Particularly, after the last two cycles of refinement were carried out we got a vastly improved fit in the high angle region (70-120° 2θ) with corresponding improvements in the various R parameters. In total, 470-500 (hkl) reflections were used for the refinement. After achieving the final fit, the C2/m unit cell was transformed into the Cm unit cell, which was used to check the goodness of fit and R parameters for the Cm case without refining any fitting parameter apart from the scale factor. We obtained the same goodness of fit and R parameters as those for the C2/m. However, further improvement in the fitting parameters could be achieved only by introducing random distortions in the oxygen atom positions in the Cm unit cell model, which were clearly unphysical. Thus the true spacegroup of the $\beta'$-$Cu_xV_2O_5$ is found to be C2/m as it describes the structure with the highest symmetry and smallest possible asymmetric unit cell. The refined cell parameters (in nm), occupancies for the Cu atom (g), the goodness of fit (S), R parameters and Durbin-Watson statistics are given in table.I. The occupancies for the rest of the atoms did not vary significantly from 1.00 and hence have been taken to be equal to 1.00. Fractional atom coordinates (x,y,z) and the constrained isotropic atomic displacement parameter, ($U_{iso}$ ($nm^2$)), obtained with C2/m spacegroup for all the investigated compositions are summarized in Table.II. The unit cell model modified as above contains 12 atoms in the asymmetric unit cell and six formula units per unit cell (total number of 50 atoms per unit cell). The typical calculated patterns, for x=0.33 and 0.60, with the corresponding experimental patterns and the differences between the two are shown in Fig.3a and Fig.3b respectively for the modified C2/m representation. The insets in both the figures show the fit obtained in the 2θ = 70º – 120º range, which is not visible clearly in the full range plot, along with the background and the difference curves. It is seen that the fit in the higher 2θ range is also quite good. All the results that are quoted in this paper have been obtained using the modified C2/m unit cell model.

|  | x=0.33 | x=0.40 | x=0.50 | x=0.55 | x=0.60 |
|---|---|---|---|---|---|
| Atom | x, y, z, Uiso (±002) and (±007) | x, y, z, Uiso (±003) and (±008) | x, y, z, Uiso (±003) and (±008) | x, y, z, Uiso (±003) and (±008) | x, y, z, Uiso (±003) and (±009) |
| Cu(8j) | 0.471,0.125,0.645,5.5 | 0.466,0.097,0.644,3.0 | 0.470,0.077,0.646,3.6 | 0.466,0.124,0.648,4.6 | 0.464,0.095,0.650,6.4 |
| V1(4i) | 0.336,0.0,0.084,9.1 | 0.334,0.0,0.091,8.2 | 0.332,0.0,0.088,10.4 | 0.332,0.0,0.086,9.1 | 0.330,0.0,0.083,9.2 |
| V2(4i) | 0.114,0.0,0.122,9.1 | 0.117,0.0,0.119,8.2 | 0.117,0.0,0.122,10.4 | 0.113,0.0,0.120,9.1 | 0.115,0.0,0.117,9.2 |
| V3(4i) | 0.284,0.0,0.403,9.1 | 0.281,0.0,0.405,8.2 | 0.287,0.0,0.408,10.4 | 0.286,0.0,0.399,9.1 | 0.282,0.0,0.408,9.2 |
| O1(2a) | 0.0,0.0,0.0,2.4 | 0.0,0.0,0.0,5.6 | 0.0,0.0,0.0,5.4 | 0.0,0.0,0.0,5.4 | 0.0,0.0,0.0,5.2 |
| O2(4i) | 0.176,0.0,0.908,2.4 | 0.185,0.0,0.935,5.6 | 0.193,0.0,0.936,5.4 | 0.186,0.0,0.920,5.4 | 0.187,0.0,0.912,5.2 |
| O3(4i) | 0.371,0.0,0.904,2.4 | 0.363,0.0,0.912,5.6 | 0.366,0.0,0.913,5.4 | 0.369,0.0,0.913,5.4 | 0.371,0.0,0.913,5.2 |
| O4(4i) | 0.442,0.0,0.212,2.4 | 0.420,0.0,0.220,5.6 | 0.433,0.0,0.212,5.4 | 0.433,0.0,0.231,5.4 | 0.434,0.0,0.215,5.2 |
| O5(4i) | 0.279,0.0,0.234,2.4 | 0.272,0.0,0.230,5.6 | 0.272,0.0,0.226,5.4 | 0.263,0.0,0.219,5.4 | 0.271,0.0,0.223,5.2 |
| O6(4i) | 0.092,0.0,0.272,2.4 | 0.081,0.0,0.275,5.6 | 0.088,0.0,0.277,5.4 | 0.085,0.0,0.285,5.4 | 0.074,0.0,0.267,5.2 |
| O7(4i) | 0.234,0.0,0.574,2.4 | 0.236,0.0,0.572,5.6 | 0.238,0.0,0.573,5.4 | 0.233,0.0,0.572,5.4 | 0.223,0.0,0.590,5.2 |
| O8(4i) | 0.415,0.0,0.496,2.4 | 0.403,0.0,0.476,5.6 | 0.397,0.0,0.470,5.4 | 0.415,0.0,0.466,5.4 | 0.413,0.0,0.481,5.2 |

Table.II: The refined fractional atom positions (x,y,z) and isotropic atomic displacement parameters ($U_{iso}$) for various Cu concentrations in $Cu_xV_2O_5$. $U_{iso}$ is in ($10^{-4} \times nm^2$) units.

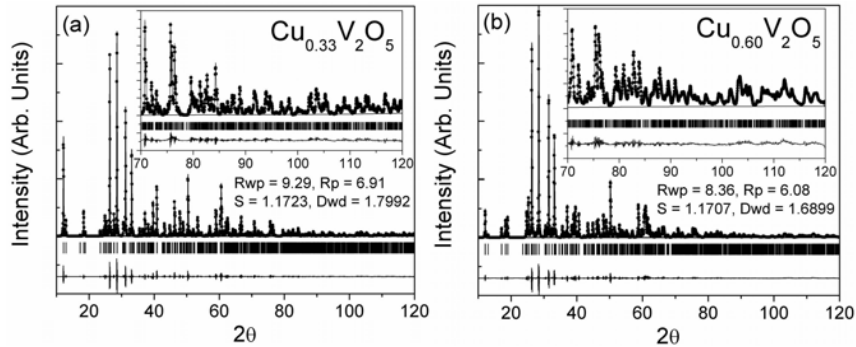

Fig.3. Observed X-ray diffraction pattern (filled circles) for (a) x=0.33 and (b) x=0.60, along with the calculated pattern (solid lines). Their difference and theoretically generated reflections (vertical lines) are shown at the bottom of the figure. Insets show fit in higher 2θ range along with background and difference.

No discontinuities are seen in the cell parameters or cell volume and all the cell parameters exhibit monotonous behavior. While a and c are observed to decrease marginally with x, b is observed to increase. β is observed to decrease with increasing x, and the volume of the unit cell remains practically constant. This behavior is similar to that observed by Kanke et al., [32]. As discussed above, the crystal structure of the $Cu_xV_2O_5$ compounds is able to accommodate a large amount of Cu without undergoing structural phase transformation. Bond lengths and angles between the various atoms can be calculated from the lattice parameters and fractional atom coordinates given in Tables I and II and compared with those given in reference 9. We have examined the V-O bond lengths and a progressive disorder in the $VO_6$ octahedra with increasing Cu concentration is observed. This basically results in a change in its orientation with respect to the b-axis [8,9]. It results in a decrease in monoclinic angle by ~0.85° and relatively large displacements in the various oxygen positions in the unit cell as compared to the V and Cu positions. The fitted values of the occupancies for all the atoms give the composition of the samples. The Cu composition of the samples as determined from the refined occupancy parameters (g) and the number of formula units per unit cell was close to the expected starting composition. It was 0.32, 0.42, 0.48, 0.52 and 0.61 for x = 0.33, 0.40, 0.50, 0.55 and 0.60 respectively. The corresponding x-ray densities were calculated to be 3736, 3887, 3913, 4015 and 4097 kg/m$^3$, which compare well with the quoted densities [9].

Maximum entropy map of electron density is obtained using the PRIMA package developed by Dilanian and Izumi [28], which maximizes the information theoretical entropy, S, subject to the constraints of the diffraction data. S is given by

$$S = -\sum_{k=1}^{N} \rho_k \ln\left(\frac{\rho_k}{\tau_k}\right), \text{ and } \rho_k = \rho_k^* / \sum_{k=1}^{N} \rho_k^* \qquad (1)$$

Where, $\rho_k$ is the normalized density at the position $r_k$ in the 3D-gridded space, $\tau_k$ the normalized density at $r_k$ derived from the previous iteration and N the total number of pixels in the unit cell; $N = N_aN_bN_c$, where $N_a$, $N_b$ and $N_c$ are the number of pixels along the a, b and c axis respectively. $\rho_k$ is obtained from the actual electron density $\rho_k^*$ at $r_k$. For the first iteration, $\tau_k = 1/N$. S is maximized subject to the following three constraints:

$$\rho_k > 0, \ C_F = \sum_{j=1}^{M_F} \frac{|F_c(h_j) - F_0(h_j)|^2}{\sigma_j^2} = M_F \text{ and } C_N = \sum_{k=1}^{N} \rho_k = 1 \qquad (2)$$

Where, $M_F$ is the total number of reflections with known phases $\alpha_j$. $F_c(h_j)$, $F_o(h_j)$ the calculated and observed structure factors for reflection $h_j$ respectively and $\sigma_j$ the estimated standard deviation (e.s.d.) of $F_o(h_j)$. All these quantities are input using the information derived from the Rietveld refinement procedure. $C_F$ is the so-called F-constraint and $C_N$ the normalization constraint. $F_c(h_j)$ is given by

$$F_c(h_j) = T \sum_{k=1}^{N} \rho_k \exp(2\pi i h_j r_k) \qquad (3)$$

where, T is the total number of electrons in the unit cell. We have ignored the so-called G-constraints for grouped reflections. This means that all the reflections that differ by Δd=0.001 or more are treated as independent reflections. Thus all the reflections calculated from the Rietveld refinement cannot be used for MEM analysis. Therefore the value of Δd is chosen such that we are able to obtain the structure factors for a maximum number of reflections for input to the maximum entropy program. The structure factors are corrected for the Lorentz polarization effects, absorption effects and anomalous scattering effects in RIETAN itself before being used in PRIMA package. It should also be mentioned here that the accuracy and precision of integrated intensities for reflections with large d-spacings should be as high as possible in MEM analysis as the information about chemical bonding is contained mainly in these reflections. In this formulation the $\rho_k > 0$ condition is automatically ensured. Since the equation for $\rho_k$ cannot be solved analytically, $0^{th}$ order single-pixel approximation has been used to calculate the MEM maps numerically. This requires a starting input electron density, which is provided by assuming an unbiased *apriori* uniform electron density in the unit cell of volume V, $\rho_k^{*0} = T/V$.

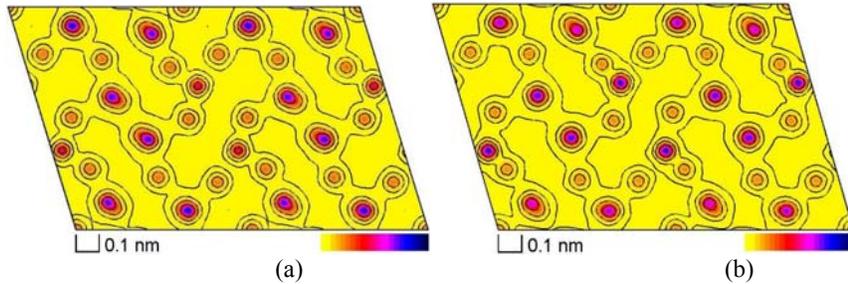

Fig.4. Electron density maps of the full unit cell projected on to (010) plane for (a) x=0.33, (b) x=0.60. All the atoms indicated by labels in Fig.1b can be located in these maps.

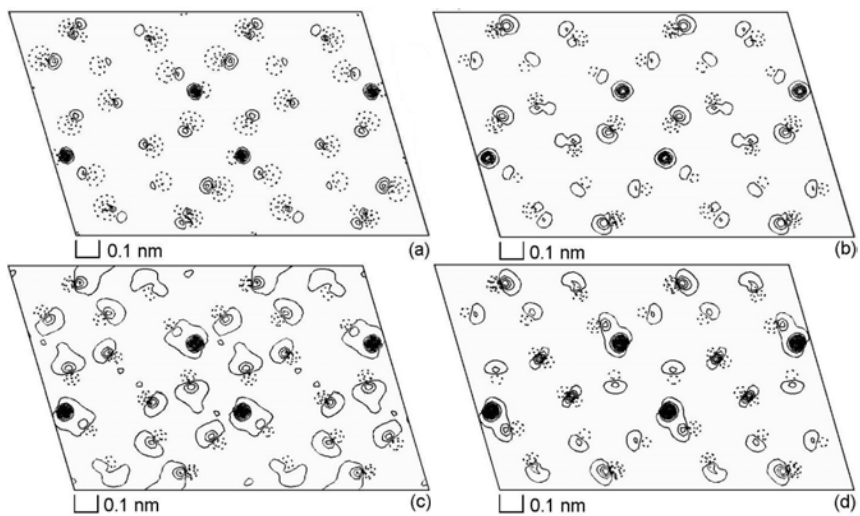

Fig.5. Difference projections maps of electron density in the (010) plane for (a) x=0.40 & 0.33, (b) x=0.50 & 0.33, (c) x=0.55 & 0.33, (d) x=0.60 & 0.33. Dotted lines denote negative electron density and solid line denote positive electron density. The positional and charge disorders induced can be clearly seen.

In the present MEM calculations, the unit cell was divided into 80 × 32 × 80 pixels along the three crystallographic directions. The resolution of MEM maps is estimated to be 0.019 nm × 0.011 nm × 0.013 nm along the a, b and c axes respectively for the x = 0.4 compound. This type of resolution is necessarily required for obtaining electron density maps where the atom positions can be observed clearly [17-25,30]. The initial value of the Lagrange multiplier, λ was set at 0.02 for all calculations and during the subsequent iterations the program automatically adjusted it. The structure factor data, corrected for Lorentz polarization, absorption effects and anomalous scattering effects were extracted from the Rietveld refinement procedure and used as input for the MEM analysis. Approximately 430-450 reflections were used for the analysis in all the samples. This reduction in total number of usable reflections is because the grouped reflections have been ignored in our calculations. The data from x=0.40 specimen were associated with maximum e.s.d. in the structure factors. The e.s.d.'s were found to be between 0.06 and 1.3 for all the reflections in all the samples under study. The reliability factors were obtained as the convergence criteria was attained after ~2000 iterations for all the samples. After the completion of REMEDY cycles, the final values of reliability factors ($R_F$ and $R_{wF}$) obtained from the MEM analysis of all the samples were 0.030 and 0.024 for x = 0.33, 0.026 and 0.023 for x = 0.40, 0.028 and 0.023 for x = 0.50, 0.028 and 0.023 for x = 0.55 and 0.026 and 0.022 for x = 0.6 respectively. The low values of $R_F$ and $R_{wF}$ in MEM analysis show the independence of the electron density patterns from the model used in the present work. Thus, the electron density maps so computed are reliable.

Projection contour maps of the MEM electron density for the full unit cell along the b-axis projected onto the (010) plane are shown in Fig.4 for (a) x = 0.33, (b) x = 0.60. All the MEM maps have been drawn in the range 0.010 – 1.230 e/($10^{-3}$nm$^3$) on a logarithmic scale. A comparison of all the figures shows that during the REMEDY cycles details of the basic structure were retained as all the atoms shown in the model in Fig.1(b) can be located on the maps. The bonding nature can be clearly seen to be covalent. Because the local difference between the MEM maps is not too evident in the figures 4, we have subtracted the 3D electron density data of x=0.33 from that of all the other Cu concentrations. This helps greatly in visualizing the local disorder induced by the increasing Cu concentration in the $V_2O_5$ sublattice. These plots are shown in Fig.5 as the difference between (a) x=0.40 and x=0.33, (b) x=0.50 and x=0.33, (c) x=0.55 and x=0.33 and (d) x=0.60 and x=0.33. As the atom positions in any two maps are not exactly same, a dumbbell shape comprising negative and positive electron density is present at each atom position. The dotted contours for negative electron density represent the atom positions in x=0.33 and the solid contours for positive electron density represent the atom positions in the corresponding Cu concentration from which x=0.33 has been subtracted. We can clearly see charge and positional disorder that is induced locally with respect to x=0.33 from these difference maps. The difference is observed to increase at the V3 atom site much more strongly that that at the V1 atom site, whereas the difference at the V2 atom site is almost constant. Numerical information can be obtained by analyzing the magnitude of electron density at the atom sites in the unmodified MEM maps. The structure dictates that the electron transfer from copper to vanadium atoms has to be mediated by the intermediary oxygen atoms. The O4, O6 and O8 atoms can mediate the electron transfer from the Cu atom to the V1, V2 and V3 atoms respectively. We have studied the changes in the electron density distribution at the Cu, O4, V1, O6, V2, O8 and V3 and O5 atomic sites. The probability of electron transfer to V1 or V3 sites is much more than that of transfer to V2 site as the distances between Cu and O4 or O8 oxygen atoms are much smaller than those between Cu and O6 atom. The electron density at the various atom sites is shown in Fig.6 for all the samples. The electron density can be seen to increase at the V1, V3 and Cu site while that at the V2 and oxygen sites remains almost unaltered. As expected, the rate of increase of electron density is the maximum at the Cu site. The electron concentration at all the oxygen sites does not change much implying oxygen atoms only act as mediator for the electron transfer from Cu to vanadium atoms. The rate of increase in the electron density at the V1 and V3 sites is almost the same. This is ~2/3 of the rate of increase in electron density at the Cu sites and 3 times the rate of increase at the V2 site. This represents a sort of average behavior of the electrons in the $V_2O_5$ sublattice and the observations can be explained on the basis of the model proposed by Goodenough [6]. From the Rietveld analysis we observe that the distance between V1–V3 is 0.3542 nm for x=0.33, which decreases to 0.3419 nm for x=0.60. Similarly, the V1-V1' (V1' is the atom in the layer directly below or above the V1 atom) distance increases from 0.3621 nm for x=0.33 to 0. 3633 nm for x=0.60. Although the V1-V2 distance is much smaller, than the V1-V3 or V1-V1' distance, the electron finds it difficult to hop along this path because of the absence of any oxygen intermediary. Thus the path for electron hopping is via V3 sub array intermediary rather than hopping from V1 to V1' atom since V1-V1' separation is much larger than V1-V3 distance [6].

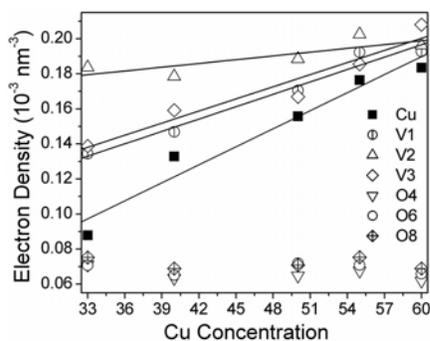

Fig.6. Plots of the variation in electron density per unit volume at various atom positions with increasing Cu concentration. The linear fits are plotted for the Cu and V atom sites as a guide to the eye only. We have studied the transfer of electrons from the Cu site to the V1, V2 and V3 sites mediated by the O4, O6 and O8 ions respectively.

## 4. Conclusions

Single phase copper vanadium oxides with a wide compositional range ($0.33 \leq x \leq 0.60$) in the $\beta'$-$Cu_xV_2O_5$ phase have been prepared. Copper intercalated into an insulating $V_2O_5$ matrix donates electrons to the vanadium atoms. Rietveld refinement of x-ray data has been carried out on these samples and good fits obtained. Lattice parameters and fractional atom coordinates deduced from the refinement indicate that the disorder in the system increases with addition of Cu. Maximum entropy method was used to derive electron density maps as a function of copper composition. The difference electron maps show that addition of copper induces positional as well as charge disorders in the V-O network. The electron hopping takes place from V1 via the V3 sub array as reported by Goodenough in the case of $\beta'$-$Cu_xV_2O_5$, which is clearly evident from the projection electron maps.

## References
[1] Wadsley AD, Acta Cryst. 1955;8:695
[2] Kanai Y, Kagoshima S and Nagasawa H, Synthetic Metals 1984;9:369
[3] Chakraverty BK, Sienko MJ and Bonnerot J, Phy. Rev. 1978;B17:3781
[4] Kapustkin VK, Volkov VL and Fotiev AA, J. Solid State Chem. 1976;19:359
[5] Casalot A and Hagenmuller P, J. Phys. Chem. Solids 1969;30:1341
[6] Goodenough JB, J. Solid State Chem. 1970;1:349
[7] Galy J, Darriet J, Casalot A and Goodenough JB, J. Solid State Chem.1970;1:339; Takayama-Muromachi E and Kato K, J. Solid State Chem. 1987;71:274

[8] Galy J, Lavaud D, Casalot A and Hagenmuller P, J. Solid State Chem. 1970;2:531
[9] Savariault JM, Deramond E and Galy J, Z. Kristall. 1994;209:405;
[10] Kato K, Takayama-Muromachi E and Kanke Y, Acta Cryst. 1989;C45:1845
[11] Mori T, Kobayashi A, Sasaki Y, Ohshima K, Suzuki M and Kobayahi H, Solid State Commun. 1981;39:1311
[12] Sperlich G, Laze WD and Bang G, Solid State Commun.1975;16:489
[13] Isobe M and Ueda Y, J. Phys. Soc. Jpn. 1996;65:1178
[14] Yamada M and Ueda Y, J. Phys. Soc. Jpn. 2000;69:1437
[15] Gilmore CJ, Acta Cryst. 1996;A52:561
[16] Sakata M and Sato M, Acta Cryst. 1990;A46:263
[17] Gull SF, In: Maximum Entropy and Bayesian Methods, Skilling J, editor, Kluwer Academic Publishers, The Netherlands, 1989.
[18] Takata M, Sakata M, Kumazawa S, Larsen F and Iversen B, Acta Cryst. 1994;A50:330
[19] Papoular RJ and Cox DE, Europhys. Lett. 1995;32:337
[20] Sivia DS and David WIF, Acta Cryst. 1994;A50:703
[21] Papoular RJ, Zheludev A, Ressouche E and Schweizer J, Acta Cryst. 1995;A51:295
[22] Knorr K, Madler F and Papoular RJ, Microporous and Mesoporous Mater. 1998;21:353
[23] Takata K, Sakurai H, Takayama-Muromachi E, Izumi F, Dilanian RA and Sasaki T, Nature 2003;422:53
[24] Dong W and Gilmore CJ, Acta Cryst. 1998;A54**:**438
[25] Takata M, Nishibori E, Umeda B, Sakata M, Yamamoto E and Shinohara H, Phys. Rev. Lett. 1997;78:3330
[26] Cheng Dong, J. Appl. Cryst. 1999;32:38
[27] Izumi F and Ikeda T, Mater. Sci. Forum 2000;321-324:198
[28] Izumi F and Dilanian RA, In: Recent Research Developments in Physics, vol. 3, Part II, Trivandrum: Transworld Research Network, 2002, p.699
[29] Izumi F, The Rigaku Journal 2000;17:34
[30] Claridge JB, Kubozono Y and Rosseinski MJ, Chem. Mater. 2003;15:1830
[31] Young RA, Editor, The Rietveld Method, IUCr and Oxford University Press, 1996
[32] Kanke Y, Takayama-Muromachi E and Kato K, J. Solid State Chem. 1989;83:69